\documentstyle[12pt,epsf]{article}
\begin{document}

\newcommand{\tstrut}{\vline height11pt depth7pt width0pt}
\def\op{{\cal O}}
\def\lsim{\mathrel{\lower4pt\hbox{$\sim$}}\hskip-12pt\raise1.6pt\hbox{$<$}\;
}
\def\Dd{\psi}
\def\pp{\lambda}
\def\ket{\rangle}
\def\BAR{\bar}
\def\xba{\overline}
\def\fa{{\cal A}}
\def\fm{{\cal M}}
\def\fl{{\cal L}}
\def\x{{\cal M}}
\def\ufs{\Upsilon(5S)}
\def\gsim{\mathrel{\lower4pt\hbox{$\sim$}}
\hskip-10pt\raise1.6pt\hbox{$>$}\;}
\def\ufour{\Upsilon(4S)}
\def\xcp{X_{CP}}
\def\ynotcp{Y}
\vspace*{-.5in}
\def\etap{\eta^\prime}
\def\bfb{{\bf B}}

\def\uglu{\hskip 0pt plus 1fil
minus 1fil} \def\uglux{\hskip 0pt plus .75fil minus .75fil}

\def\slashed#1{\setbox200=\hbox{$ #1 $}
    \hbox{\box200 \hskip -\wd200 \hbox to \wd200 {\uglu $/$ \uglux}}}
\def\slpar{\slashed\partial}
\def\sla{\slashed a}
\def\slb{\slashed b}
\def\slc{\slashed c}
\def\sld{\slashed d}
\def\sle{\slashed e}
\def\slf{\slashed f}
\def\slg{\slashed g}
\def\slh{\slashed h}
\def\sli{\slashed i}
\def\slj{\slashed j}
\def\slk{\slashed k}
\def\sll{\slashed l}
\def\slm{\slashed m}
\def\sln{\slashed n}
\def\slo{\slashed o}
\def\slp{\slashed p}
\def\slq{\slashed q}
\def\slr{\slashed r}
\def\sls{\slashed s}
\def\slt{\slashed t}
\def\slu{\slashed u}
\def\slv{\slashed v}
\def\slw{\slashed w}
\def\slx{\slashed x}
\def\sly{\slashed y}
\def\slz{\slashed z}

\rightline{AMES-HET 01-11}
\rightline{BNL-HET-01/25}
\rightline{FERMILAB-Pub-01/177-T}

\begin{center}
%
%
%
%
%
{\large\bf
Determining the phases $\alpha$ and $\gamma$ from time-dependent CP
violation in $B^0$ decays to $\rho$($\omega$) + pseudoscalar
}
\vspace{.2in}

David Atwood$^{1}$\\
\noindent Dept. of Physics and Astronomy, Iowa State University, Ames,
IA
50011\\
\medskip

Amarjit Soni$^{2}$\\
\noindent Theory Group, Brookhaven National Laboratory, Upton, NY
11973\\
\footnotetext[1]{email: atwood@iastate.edu}
\footnotetext[2]{email: soni@bnl.gov}
\end{center}
\vspace{.15in}

\begin{quote} 

{\bf Abstract:} 

A method is proposed for the determination of the unitarity angle
$\alpha$ through tree penguin interference.  The modes needed would be of
the form $B^0/\xba B^0\to \rho^0 \x$ and $B^0/\xba B^0\to \omega \x$ where
$\x$ is spin-0 $u\xba u/d\xba d$ meson, for instance $\x=\pi^0$, $\eta$,
$\eta^\prime$, $a_0$ or $f_0$. An analogous method can also determine
$\gamma$ using $\x=K_S$ or $K_L$.  The validity of the theoretical
approximations used may be tested by over determining $\alpha$ with
several modes.  If two or more modes are used, the determination has a
four-fold ambiguity but additional information from pure penguin decays or
theoretical estimates may be used to reduce the ambiguity to $\alpha$,
$\alpha+\pi$.  The method as applied to determining $\gamma$ is probably
less promising.

\end{quote}


The early indications of CP violation~\cite{cdf_ref,belle_ref,babar_ref}
in the neutral B system is an important development in our understanding of
this phenomenon.  It is expected that in the near future the angle $\beta$
of the unitarity triangle will be determined with considerable accuracy.  
This is a crucial first step in the program of verifying that the
Cabibbo-Kobayashi-Maskawa (CKM) matrix~\cite{cabibbo} of the Standard
Model is the origin of CP-violation in the $B$ and $K$ systems. To
complete this task, the more difficult angles $\alpha$ and $\gamma$ also
need to be determined. Information about these angles, when combined with
the information concerning the sides of the CKM
triangle~\cite{CKMparadgim}, will either provide an impressive
verification for the CKM paradigm or convincing evidence for the presence
for new physics.

In this Letter we propose a method to extract $\alpha$ through the
interference of the $b\to u\xba u d$ tree with the $b\to d$ penguin;  an
analogous method may also be used for $\gamma$ through the interference of
the $b\to u\xba u s$ tree with the $b\to s$ penguin. In particular, we
will require time dependent observations of $B^0/\xba B^0\to \rho \x$ and
$B^0/\xba B^0\to \omega \x$ where $\x$ is any self-conjugate spin-0 state
with a suitable quark content. For instance to obtain $\alpha$ we can use
$\x=\pi^0$, $\eta$, $\eta^\prime$, $a_0$ or $f_0$ as well as related
excited states, while to obtain $\gamma$ we may use $\x=K_S$, $K_L$ or
related spin 0 kaonic resonances.  Comparison of $\rho \x$ and $\omega \x$
data for each (pseudo)-scalar $\x$ gives up to a 4-fold degeneracy in the
determination of $\sin 2\alpha$ (or $\sin 2\gamma$) and when information
from two or more modes is combined, a single value of this quantity should
emerge.  The remaining four fold degeneracy in $\alpha$ ($\gamma$) can
then be reduced to two-fold by either using theoretical information
concerning the tree or penguin amplitudes or by measuring a pure penguin
mode related by SU(3).

There have been several methods proposed to extract these angles. For
$\alpha$ one can consider oscillation effects in $B^0\to\pi^+\pi^-$
although one must account for the penguin through isospin
analysis~\cite{alpha1} by observing $B^0\to\pi^0\pi^0$.  Since the
branching ratio to $\pi^0\pi^0$ is expected to be small and hard to
observe, it may be preferable to consider three $\pi$ final states.  
In~\cite{alpha2}, $\alpha$ is determined through isospin analysis of
$B^\pm\to\pi^\pm\pi^+\pi^-$, $B^\pm\to\pi^\pm\pi^0\pi^0$ and $B^0/\xba
B^0\to\pi^+\pi^-\pi^0$. In this approach one can take advantage of
resonance effects in the Dalitz plot; however, there may be problems in
precise modeling of the resonance structure. Another method for extracting
$\alpha$ from the interference of $u$-penguins with $t$-penguins in
$B^0\to K^{(*)}K^{(*)}$ may overcome the disadvantages of the $2\pi$ and
$3\pi$ final states~\cite{londonKK} although the analogous $B_s$ decays
are required for the analysis. The angle $\gamma$ may be extracted via
direct CP violation through the interference of $b\to u\xba c s$ and $b\to
c\xba u s$~\cite{adsglw} and via time-dependent studies
of CP violation in the $B_s$ system~\cite{bs_time}.

In the method discussed here, we rely on the following two approximations:

\begin{itemize}

\item[(1)] The contribution of the electro-weak penguin (EWP) is small.

\item[(2)] The $q\bar q$ pair which arises in a strong penguin does not
form the $\omega$ in the final state.

\end{itemize}

\noindent Recall that the EWP are also assumed to be small in some
of the other
proposed methods~\cite{alpha1,alpha2} mentioned in the preceding
paragraph for extracting $\alpha$ since they rely on isospin. Our
second assumption is true at lowest order in perturbation theory by color
conservation and, as discussed in~\cite{last_paper}, using renormalization
group improved perturbation theory one can show that it is valid to a few
percent in general.

Let us first consider the case of determining $\alpha$ through the
interference of the $b\to d$ penguin with the $b\to u\xba u d$ tree. For a
given final state of this form $f$ let us define $A$ to be the amplitude
for $B^0\to f$ and $\xba A$ the amplitude for $\xba B^0\to f$. The SM
amplitude for $B^0\to f$ can be written as:

\begin{eqnarray}
A_f = T(f) v_u + P_u(f) v_u + P_c(f) v_c + P_t(f) v_t
\equiv \hat T_f v_u + \hat P_f v_t
\label{hat-def}
\end{eqnarray}

\noindent where $T(f)$ is the tree contribution to the amplitude, $P_i(f)$
is the penguin contribution due to the diagram with an internal quark of
type $u_i$ and $v_i=V^*_{ib}V_{id}$.  Unitarity of the CKM matrix allows
us to express the amplitudes in terms of $\hat T_f=T(f)+P_u(f)-P_c(f)$
and $\hat P_f=P_t(f)-P_c(f)$.

If we denote 
$t_f=|\hat T_f v_u|$ 
and
$p_f=|\hat P_f v_t|$ 
then, using the phase convention
of~\cite{wolfenstein_conv},
we can write:

\begin{eqnarray}
A_f&=&		t_fe^{+i\gamma}+p_fe^{-i\beta}
\nonumber\\
\xba A_f&=&	t_fe^{-i\gamma}+p_fe^{+i\beta}
\label{tp-split}
\end{eqnarray}

\noindent
In the limit that $\Delta \Gamma/\Gamma$ is small, which is 
a good approximation  
for the
$B^0$ meson, the time dependent decay rate is:

\begin{eqnarray}
{1\over \Gamma_{B^0}}
{d\Gamma(f)\over d\tau}
=
{1\over 2}e^{-|\tau|}
\left (
X_f 
+b Y_f\cos{x_b\tau}
-b Z^I_f\sin{x_b\tau}
\right )
\label{time_series}
\end{eqnarray}

\noindent
where $b=+1$ (-1) for  $B^0$ ($\xba B^0$)-mesons, $\tau=\Gamma_B t$ and
the coefficients 
$X_f$, $Y_f$ and $Z^I_f$
are related to the amplitudes 
in eq.~(\ref{tp-split}) by:

\begin{eqnarray}
X_f&=&(|A_f|^2+|\xba A_f|^2)/2
\nonumber\\
Y_f&=&(|A_f|^2-|\xba A_f|^2)/2
\nonumber\\
Z^I_f&=&Im(e^{-2i\beta} A_f^*\xba A_f)
\label{obs_def1}
\end{eqnarray}

\nonumber
From these, we can also determine, up to a two fold ambiguity,
the quantity:

\begin{eqnarray}
Z^R_f=Re(e^{-2i\beta} A_f^*\xba A_f)
\label{obs_def2}
\end{eqnarray}

\noindent since $(Z^I_f)^2 +(Z^R_f)^2 =X_f^2-Y_f^2$.
We can expand $X$
and $Z$ in terms of the tree and penguin amplitudes as:

\begin{eqnarray}
X_f&=& 		|p_f|^2 + |t_f|^2 - 2r_f\cos\alpha
\nonumber\\
Z^R_f&=& 	|p_f|^2 + |t_f|^2\cos 2\alpha - 2r_f\cos\alpha
\nonumber\\
Z^I_f&=&	|t_f|^2\sin 2\alpha - 2r_f \sin\alpha
\label{key_system}
\end{eqnarray}

\noindent
where $r_f=Re(t_f p^*_f)$.

If we eliminate $|t_f|^2$ and $r_f$ from eq.~(\ref{key_system})
and
denote $Z_f=Z^R_f+iZ^I_F=e^{-2i\beta}A_f^*\xba A_f$, 
we
obtain:

\begin{eqnarray}
X_f-Z^R_f\cos2\alpha-Z_I\sin 2\alpha
\equiv
X_f-Re(Z_f e^{-2i\alpha})
=(1-\cos 2\alpha) |p_f|^2
\label{p-eqn}
\end{eqnarray}

\noindent
Likewise, if we eliminate $|p_f|^2$ and $r_f$ we obtain:

\begin{eqnarray}
X_f-Z^R_f=(1-\cos 2\alpha)|t_f|^2
\label{t-eqn}
\end{eqnarray}

Our primary approach to obtaining $\alpha$ is to use eqn.~(\ref{p-eqn}) to
relate two modes by matching $B^0\to \rho^0 \x$ and $B^0\to\omega \x$,
$\x=$ any spin-0 $u\xba u/d\xba d$ meson. The fact that the magnitude of
the penguin is the same for these two modes in principle provides enough
equations to cleanly determine $\alpha$. This, however, will normally
result in discrete ambiguities, so it is important to supplement this
method in such a way as to over determine $\alpha$ and thus reduce these
ambiguities. This may be accomplished in a number of ways:

\begin{enumerate}

\item[(a)] Use the same procedure with a number of different candidates
for $\x$, indeed $\x=\{\pi^0$, $\eta$, $\eta^\prime$, $a_0$, $f_0\}$ and
higher excited states may all be viable. In addition a meson with non-zero
spin may be used for $\x$ if one analyzes the angular distributions of the
final state to determine the time dependent magnitude of the specific
helicity amplitudes. Of particular interest in this category are
$f=\rho^0\rho^0$, $\rho^0\omega$ and $\omega\omega$ all three of which
share a common $|p_f|^2$.

\item[(b)] Use SU(3) to determine the magnitude of the penguin amplitude
from a pure penguin process. As we shall see, even if there are
considerable theoretical errors in the application of SU(3), this
additional information will often resolve the ambiguity between $\alpha$
and $\pi-\alpha$ which cannot be resolved purely from the above
$\rho-\omega$ matching.

\item[(c)] Other theoretical bounds such as the ratio between the tree and
the penguin, again even with the presence of appreciable uncertainty, may
be used to constrain the results providing a means to distinguish between
the different ambiguous solutions.

\item[(d)] Finally, 
this method
may be used in conjunction with other methods for
obtaining $\alpha$. In the Standard Model all such methods should agree on
a single value for $\alpha$ and so all independent methods to determine
$\alpha$ should be used in parallel.

\end{enumerate}

\noindent We now illustrate our approach to extracting $\alpha$ from a
comparison of $\rho^0 \x$ with $\omega \x$ for the specific case when
$\x=\pi^0$.

It follows from approximations (1) and (2) then that the penguin
contributes to these final states only via the $d\xba d d\xba d$ channel,
therefore $p_{\rho^0\pi^0}=-p_{\omega\pi^0}$ and thus from
eqn.~(\ref{p-eqn}):

\begin{eqnarray}
X_{\rho^0\pi^0}-X_{\omega\pi^0}
=
Re(e^{-2i\alpha}(Z_{\rho^0\pi^0}-Z_{\omega\pi^0}))
\end{eqnarray}

\noindent If we then define

\begin{eqnarray}
\xi=X_{\rho^0\pi^0}-X_{\omega\pi^0};
~~~~~~~
\zeta e^{i\theta}=Z_{\rho^0\pi^0}-Z_{\omega\pi^0}
\end{eqnarray}

\noindent we obtain

\begin{eqnarray}
\alpha=
(\theta\pm\cos^{-1}{\xi\over\zeta})/2 
+\left \{ 
\begin{array}{l}
0\\
or\\
\pi
\end{array}
\right.
\label{alpha_solve}
\end{eqnarray}

\noindent In general, this will give a 16-fold ambiguity in $\alpha$
caused by the four fold ambiguity in eqn.~(\ref{alpha_solve}) together
with the sign ambiguities in $Z^R_{\rho^0\pi^0}$ and $Z^R_{\omega\pi^0}$.
However, if $|\xi/\zeta|>1$ for some choices of the signs for $Z^R_f$,
there will be no corresponding solutions for $\alpha$ and in such cases
the degeneracy will be smaller.  In addition, notice that if one takes the
incorrect sign choice for both $Z^R_{\rho^0\pi^0}$ and $Z^R_{\omega\pi^0}$
and also takes the incorrect sign choice for $\cos^{-1}(\xi/\zeta)$, one
will obtain $\pi/2-\alpha$.  There will therefore always be a four fold
ambiguity in this method with $\alpha$, $\alpha+\pi$, $\pi/2-\alpha$ and
$3\pi/2-\alpha$ being possible solutions. Note that all these solutions
have the same value of $\sin 2\alpha$ so, in effect, this method
determines the value of $sin2\alpha$ with up to a 4-fold ambiguity.

The $\rho-\omega$ comparison is clean in that no theoretical assumptions
are required beyond (1) and (2) so the determination of $\sin2\alpha$ is
limited only by statistics. Furthermore it is self-contained in that only
information from the $B^0$ is needed, therefore it may be used at
asymmetric B-factories or at hadronic $B$ facilities although the ability
to detect $\pi^0$ is required in the modes we consider.

As suggested above, the first step to resolve the ambiguity is to apply
eqn.~(\ref{alpha_solve}) to as many final states as possible. By combining
the results from several values for $\x$, one should be able to extract a
unique value for $\sin 2\alpha$ (hence a four-fold ambiguity for
$\alpha$).

One way to further reduce this ambiguity to 2-fold is to use SU(3) to
estimate the magnitude of $p_f$ from a related pure penguin $b\to s$
transition. In the case of the $\rho^0\pi^0/\omega\pi^0$ final states, the
appropriate pure penguin decay is $B_s\to K_S K^*_S$ or $K_LK^*_L$ where
$K^*_{S,L}$ means a neutral $K^*$ meson which decays to a final state
$\pi^0 K_{S,L}$.  Note that in this mode the amplitude for the spectator
to form the pseudo-scalar and the amplitude for the spectator to form the
vector states are combined in the same way as they are in the
$B^0\to\rho^0\pi^0$ case.  This decay is a pure penguin assuming that the
rescattering contribution of tree diagrams is negligible and so its decay
rate allows us to estimate $|p_f|$. Generally the penguin amplitude
obtained by $\rho-\omega$ comparison will be different for $\alpha$ and
$\pi/2-\alpha$.

Bearing in mind the SU(3) assumption used to determine $|p_f|$ from
$B_s\to K_S K^*_S$ or $K_L K^*_L$, this pure-penguin method could also be
used as a ``stand-alone'' method for determining $\alpha$ by rearranging
eqn.~(\ref{p-eqn}) into:

\begin{eqnarray}
X_f-|p_f|^2=Re((Z_f-|p_f|^2)e^{-2i\alpha})
\Rightarrow
\alpha
=(\mu_f\pm \cos^{-1}{ X_f-|p_f|^2\over R_f})/2
+\left \{ 
\begin{array}{l}
0\\
or\\
\pi
\end{array}
\right.
\end{eqnarray}

\noindent 
where 
$R_f$ and $\mu_f$ are defined by
$Z_f-|p_f|^2=R_fe^{i\mu_f}$.
This gives up to an 8-fold ambiguity:  2-fold from the sign of
$Z^R_f$, two fold from the sign of $\cos^{-1}$, and two fold from the
$mod~\pi$ ambiguity. In this context both $f=\rho\pi^0$ and
$f=\omega\pi^0$ may be used so the ambiguity of this stand-alone method
may, in principle, be reduced to the 2-fold $mod~\pi$ ambiguity.

Theoretical input can also be used to reduce the ambiguities and refine
the determination of $\alpha$. Estimates of the penguin or tree amplitudes
may serve this purpose.  In this context it is useful to note that the
ratio of eqn.~(\ref{p-eqn}) with eqn.~(\ref{t-eqn})  gives us the
tree-penguin ratio as a function of $\alpha$:

\begin{eqnarray}
\left |{t_f\over p_f}\right |^2
=
{
X_f-Z_f^R
\over
X_f-Z_f^R\cos 2\alpha-Z_f^I\sin 2\alpha
}
\label{rat-eqn}
\end{eqnarray}

\noindent A theoretical range for this ratio will therefore translate
directly into a range of values for $\alpha$ given the experimental inputs
for each final state $f$ separately.

%
%
%
%
%
%

The degree of precision which may be achieved depends very much how the
different ambiguous solutions happen to align up for various modes and in
turn on the phases for the various amplitudes involved. In order to
illustrate the situation, let us construct a toy model based on related
observed branching ratios taking $\x=\pi^0$ and $\eta$.  In the case of
$\x=\eta$, the analogous $b\to s$ penguin modes:  $B^-\to K^{*-}\eta$ and
$B^0\to K^{*0}\eta$ have been observed at CLEO~\cite{hepex9912059,etanote}
with an average branching ratio of $\sim 2\times 10^{-5}$.  We thus expect
that $|p_{\rho\eta}|^2\approx |V_{td}/V_{ts}|^2~Br(K^*\eta)$, (using
amplitudes in units that square to branching ratio), so taking
$|V_{td}/V_{ts}|\approx 0.2$ we obtain $|p_{\rho\eta}|^2=0.8\times
10^{-6}$. In order to estimate the penguin rate for $\rho^0\pi^0$ let us
consider the analogous pure penguin mode ($\phi K$) which has been
observed at BaBaR~\cite{hepex0105001} at a rate of $Br(B\to \phi K)\approx
10^{-5}$. As before we estimate $|p_{\rho^0\pi^0}|^2\approx 0.4\times
10^{-6}$.

The tree process in this channel which have been observed at
CLEO~\cite{hepex0006008} are $\pi^-\rho^0$ and $\pi^0\rho^-$ both of which
have a branching ratio $\sim 10^{-5}$. These are color allowed while the
processes we are interested in are color suppressed, hence we will assume
that $|t_f|^2\sim O(10^{-6})$.

For the purpose of generating specific illustrative examples, we will
assume that for each $\x$, we will parameterize the amplitudes in the
following general form:

\begin{eqnarray}
t_{\rho \x}  =A_0^\x
~~~~
t_{\omega \x}=a_\omega^\x e^{i \psi_\omega^\x} A_0^\x
~~~~
p_{\rho \x}=-p_{\omega \x}= a_p^\x e^{ i\psi_p^\x} A_0^\x
\end{eqnarray}

\noindent
where we will specifically look at the parameter set:

\begin{eqnarray}
&&A_0^\pi=A_0^\eta=10^{-3},
~~~~
a_\omega^\pi=1.1=
a_\omega^\eta=1.1,
~~~~
a_p^\pi=0.65,
~~~~
a_p^\eta=0.90,
\nonumber\\
&&\psi_\omega^\pi=-90^\circ,
~~~~
\psi_\omega^\eta=15^\circ,
~~~~
\psi_p^\pi=20^\circ,
~~~~
\psi_p^\eta=10^\circ,
\label{toy-numbers}
\end{eqnarray}

\noindent which are consistent with the estimated rates above.  Further,
for the purpose of this illustration, we will also assume that the true
value of $\alpha=75^\circ$.

Given perfect experimental data, the set of solutions for $\alpha$ in the
case of $\x=\pi^0$ are $\{52.4^\circ$, $75^\circ$, $75.5^\circ$,
$114.1^\circ\}$ together with angles related to these by $\alpha\to
\pi/2-\alpha$;  $~\alpha\to 3\pi/2-\alpha$ and $\alpha\to\pi+\alpha$.  In
the $\x=\eta$ case, the corresponding solution set is $\{75^\circ$,
$78.9^\circ$, $82.6^\circ$, $86.9^\circ\}$ and the related angles. Only
the true solution and the three other related 
angles 
(i.e. $15^\circ$, $195^\circ$ and $255^\circ$) 
are common to both
sets.

To get an idea of how well we can do with the statistical errors from a
finite amount of data, let us define: 

\begin{eqnarray}
\tilde N_0= 
\left [
({\rm number~of~B^0})+({\rm number~of~\xba B^0})
\right ]
\cdot ({\rm acceptance})
\end{eqnarray}

Let us now derive a chi-squared function, $\chi^2$, using
statistical errors of the input quantities $X_f$, $Y_f$ and $Z^I_f$. If we
define $N_f=(Br(B^0\to f)+Br(\xba B^0\to f)) \tilde N_0 /2$ then:

\begin{eqnarray}
(\Delta X_f)^2=X_f^2/N_f
\end{eqnarray}  
   
\noindent

\noindent
If we want to determine $Y_f$ and $Z^I_f$ from the time dependent
distributions in eq.~(\ref{time_series}),
we can use the expectation value of time dependent operators which are
proportional to these quantities. Using the optimal observable as defined
in~\cite{optobs} we obtain:

\begin{eqnarray}
&g_Y=b 
\left(
1+4x_b^2\over 1+2x_b^2
\right)
\cos(x_b\tau);
~~~~~~~
&g_{Z^I}=b 
\left(
1+4x_b^2\over 2x_b^2
\right)
\sin(x_b\tau);
\nonumber\\
&<g_Y>=Y_f/X_f;
&<g_{Z^I}>=Z^I_f/X_f
\end{eqnarray}

\noindent
Assuming a tagging efficiency $T$, the statistical errors in these
observables are:

\begin{eqnarray}
<(\Delta g_y)^2>
&=&
{1\over T N_f}
\left (
{1+4x_b^2\over 1+2 x_b^2}
-
{Y^2\over X^2}
\right )
\nonumber\\
<(\Delta g_{Z^I})^2>
&=&
{1\over T N_f}
\left (
{1+4x_b^2\over 2 x_b^2}
-
{Z_I^2\over X^2}
\right )
\label{osc-errors}
\end{eqnarray}

In Fig.~1 we show the minimum $\chi^2$ as a function of $\alpha$ for these
two cases given $\tilde N_0=10^9$ and tagging efficiency $T=0.5$ assuming
that the central values are those given by the parameters in
eq.~(\ref{toy-numbers}). The $\x=\pi^0$ case is indicated by the dashed
curve, the $\x=\eta$ case by the dotted curve and the sum by the solid
curve. By construction, in each of the cases the curve must hit 0 for
possible solutions. This is clearly true in the case of $\x=\pi^0$ while
for $\x=\eta$, the solutions occur in tight clumps so, for instance, in
the range $75^\circ$ --- $86.9^\circ$, $\chi^2$ is close to 0 but rises
rapidly outside of this range.  Looking at the sum, 1$\sigma$ uncertainty
range about $75^\circ$ is $75^{+7.4}_{-3.5}$ (as well as the corresponding
range around $15^\circ$). It is asymmetric due to the $\x=\eta$ curve.  If
we lower $\tilde N_0$ to $2\times 10^8$ (i.e. scale $\chi^2$ down by a
factor of 5), then the range becomes $75^{+15.6}_{-6.8}$ where most of the
bound now depends on the $\x=\eta$ case. Likewise for $\tilde N_0=10^8$
the range is $75^{+18.9}_{-8.8}$.  Clearly if a single mode has a tight
clump of solutions as exemplified by the $\x=\eta$ case, early data can
put relatively tight bounds on the solution but it is difficult to
distinguish between the members of the clump. To refine the solution with
more statistics a mode with more widely scattered solutions such as
$\x=\pi^0$ is helpful.

Fig.~1 also shows how a pure penguin mode may be helpful in distinguishing
between the solution at $75^\circ$ and the solution at $15^\circ$. The
magnitude of the penguin amplitude in the $\x=\pi^0$ case for the 
minimum $\chi^2$
solution at each $\alpha$ is shown with the dash-dotted curve.
This is clearly very different in the two regions illustrating how pure
penguin data from $B_s\to K_S K_S^*$ may solve this ambiguity. Of course
the entire graph repeats itself in the range $180^\circ-360^\circ$ 
leaving a $mod~\pi$ ambiguity.

Fig.~2 plots the same quantities as Fig.~1 where we have changed
$\psi^\eta_p$ to $190^\circ$ to illustrate a somewhat different behavior
for the $\x=\eta$ case. In this case, the solution set is $\{75^\circ$,
$80.1^\circ$, $101.3^\circ$, $106.2^\circ\}$ and we have somewhat better
determination of the true solution since the $\x=\eta$ mode now only has
one false solution near to the true one. The two false solutions for
$\x=\eta$ near $\sim 105^\circ$ are eliminated by the $\x=\pi^0$ data
although there is a local minimum of $\chi^2$ in that vicinity since the
latter case has a false solution at $114.1^\circ$. In particular, for
$\tilde N_0=10^9$, the 1-$\sigma$ range about the true solution is
$75^{+6.5 }_{-2.5 }$ while if $\tilde N_0=2\times 10^8$ then the range is
$75^{+9.3 }_{-4.8 }$ but the other minimum near $108^\circ$ begins to
become a viable solution.

The information from this method can and should be combined with various
other methods for determining $\alpha$. In particular the method
of~\cite{alpha2} uses one of the same modes ($\rho^0\pi^0$) so jointly
fitting $\alpha$, $t_{\rho\pi^0}$ and $p_{\rho^0\pi^0}$ between the two
data sets will produce more constrained results. In addition, since that
method uses cross-channel interference, the ambiguity between $\alpha$ and
$\alpha+\pi$ may be resolved.  The interference between $B^\pm \to
\rho^\pm \rho^0$ and $B^\pm \to \rho^\pm \omega$ in the
$\omega\to\pi^+\pi^-$ channel as discussed in~\cite{gardner} may also be
used to address this ambiguity.

The comparison between $\alpha$ as determined here and the value
determined by the method of~\cite{londonKK} is particularly interesting
since in that case $\hat T$ is purely the u-penguin; deviation between the
two values of $\alpha$ could therefore be an indication of physics beyond
the Standard Model. This relation between the u-penguin phase and the tree
phase may also be tested in the method of~\cite{alpha2} through the
comparison of charged and neutral $B$ decays.  Furthermore, the methods
in~\cite{last_paper} depend only on direct CP violation and so a
discrepancy with the methods involving time-dependent CP violation could
indicate new physics in $B\xba B$ oscillation.

To obtain the angle $\gamma$ we now consider modes which are sensitive to
the interference of the $b\to s$ penguins with the $b\to s u\xba u$ tree.
The spin-0 particle which recoils against the $\rho/\omega$ should
therefore be a $K_S$, $K_L$ or any other spin-0 kaonic resonance that is
self conjugate by decaying to a $K_S$ or $K_L$~~\cite{kaon_note} in the
final state.  For this analysis we need to assume that an accurate
knowledge of $\beta$ is available, which will likely be the case.

Let us denote such final states as $g$.
In this case, the weak phase of the tree is still $+\gamma$ but the $b\to
s$ penguin phase is $\approx 0$ in the Standard Model. The decomposition
analogous to
eqn.~(\ref{tp-split}) thus gives:

\begin{eqnarray}
A_g&=&          \tilde t_g e^{+i\gamma}+ \tilde p_g
\nonumber\\
\xba A_g&=&     \tilde t_g e^{-i\gamma}+ \tilde p_g
\end{eqnarray}

\noindent where $\tilde t_g = |\hat T_g V_{ub}^* V_{ud}|$ and $\tilde p_g
= |\hat P_g V_{tb}^* V_{td}|$.  Using the observables previously
introduced in eqns.~(\ref{obs_def1},\ref{obs_def2}) we can define

\begin{eqnarray}
\tilde Z_g= Z_g e^{+2i\beta};
~~~
\tilde Z^R_g=Re(\tilde Z_g);
~~~{\rm and}~~~
\tilde Z^I_g=Im(\tilde Z_g).
\end{eqnarray}

\noindent
where we need to know $\beta$ in order to obtain $\tilde Z$ from 
$Z$. As in the case of $\alpha$ we find that:

\begin{eqnarray}
X_g- Re(\tilde Z_g e^{+2i\gamma})=(1-\cos\gamma)|p_g|^2
\end{eqnarray}

\noindent
We can use this equation in the same way as eqn.~(\ref{p-eqn}) above since
here too the penguin produces $\rho$ and $\omega$ via the 
$d\xba d$ channel so we can equate $|p_g|^2$ between $\omega K_s$ and 
$\rho^0 K_s$.
If we now define:

\begin{eqnarray}
\tilde\zeta e^{i\tilde \theta}=\tilde Z_{\rho^0 K_s}-\tilde Z_{\omega
K_s}
\end{eqnarray}

\noindent
then

\begin{eqnarray}
\gamma=
-(\tilde \theta\pm\cos^{-1}{\xi\over\tilde \zeta})/2
~~~or~~~
\pi-(\tilde \theta\pm\cos^{-1}{\xi\over\tilde \zeta})/2
\label{gamma_solve}
\end{eqnarray}

In this case, however we are somewhat restricted in the number of modes $g$
which are experimentally accessible since spin-0 kaonic resonances are
required and, aside from $K^0$, these particles are close in mass to higher
spin resonances with similar decay modes. One could, however, gain
additional modes by time dependent angular analysis of
$g=\rho/\omega~K^{*0}$ and treating each of the three helicity states as a
separate mode.

As in the $\alpha$ case, one can also resolve the ambiguities by comparing
this to a pure penguin amplitude. In this case the pure penguin which is
appropriate is $B^\pm\to K^0\rho^\pm$. This has the advantage that it will
also be produced at asymmetric B-factories.

Using these methods for the determination of $\gamma$, however is likely
to be less promising than the $\alpha$ determination.  Although the
branching ratio to such modes is relatively large ($O(10^{-5}$)), the
color suppressed tree amplitude is only about $4\%$ of the penguin so that
the interferences effects required to solve for $\gamma$ will be $\lsim
4\%$.

To see how the method works in this case, it is useful to consider the
limit that $t_g<<p_g$, we can then use the following approximate
relations:

\begin{eqnarray}
\tan\gamma
&=&
{\tilde Z^I_g\over  |p_g|^2-X_g}
+O(|t_g/p_g|^2) 
\label{approx_gamma1}
\\
\tan\gamma
&=&-
{
\tilde Z^I_{\omega K}-\tilde Z^I_{\rho^0 K}
\over  
X_{\omega K} 
-
X_{\rho K}
}
+O(|t_g/p_g|^2) 
\label{approx_gamma2}
\end{eqnarray} 

\noindent In the first relation, if $|p_g|$ is known from the analogous
pure penguin mode, this gives $\gamma$ up to a 4-fold ambiguity (2-fold
for $Z_R\to -Z_R$ and 2-fold for $\gamma\to \gamma+\pi$).

In order to estimate the number of $B$-mesons required we note that the
key measurement would be the determination of $\tilde Z_I$ which would in
turn rely on an accurate determination of $Z_I$. Assuming that
$Z_I/X\approx 0.04$, we see from eqn.~(\ref{osc-errors}) that to get a
3-sigma determination of this quantity requires $N_g\approx(X_g/\Delta
Z^I_g)^2(1+4x_b^2)/(2 T x_b^2)$ so that if we take $X_g/\Delta
Z^I_g=3/0.04$ we obtain $N_g=3\times 10^4$. If the branching ratio for
$B^0\to\omega K^0$ is roughly the same as the measured branching
ratio~\cite{PDB} $B^+\to\omega K^+$ of $1.5^{+0.7}_{-0.6} \times 10^{-5}$,
then $\tilde N_0\approx 2\times 10^9$. With this number of B-mesons,
determination of $X_g$ and $|p_g|^2$ to a precision of $<1\%$ which is
also required should be statistically possible.  However very tight
control over the backgrounds would be necessary to accurately determine
the denominator of eqn.~(\ref{approx_gamma1}).  
Eqn.~(\ref{approx_gamma2}) can give $\gamma$ up to an 8-fold ambiguity and
the experimental requirements are similar.

In summary, our primary method for determining $\alpha$ involves observing
the time dependent decays $B^0/\xba B^0\to \rho^0 \x$ and $B^0/\xba B^0\to
\omega^0 \x$ where $\x$ is a $u\xba u/d\xba d$ spin-0 meson.  If we assume
that electro-weak penguins are negligible and that the gluon fragmentation
to an $\omega$ is also negligible, the penguin contribution to these two
modes has the same magnitude. This allows us to solve for $\alpha$ with up
to 16-fold discrete ambiguity. These ambiguities may be reduced to 4-fold
by combining two or more such modes.  Using other theoretical input or
estimating the magnitude of the penguin by SU(3) related pure penguin
modes one can reduce this to a two fold ambiguity. The precision that can
be achieved depends on the various amplitudes and strong phases involved
and it was found that reasonable bounds might begin to be achieved with
$\tilde N_0\sim 10^8$ while with $\tilde N_0=10^9$ relatively tight bounds
on $\alpha$ appear possible. In the case of $\gamma$ it is possible to use
the analog of this method for the decays $B^0/\xba B^0\to \rho^0 K_S$ and
$B^0/\xba B^0\to \omega K_S$ and in principle it can be carried out with
$\tilde N_0\sim 10^9$. In this case however the relevant interference
effects are small so that systematic errors in the measurements must be
controlled to $\lsim 1\%$.

%
%
%
%
%
%
%
%
%

\bigskip

This research was supported in part by US DOE Contract Nos.\
DE-FG02-94ER40817 (ISU) and DE-AC02-98CH10886 (BNL)


\newpage

\begin{figure}
\caption{
The $\chi^2$ function for the minimum $\chi^2$ solution at various
values of $\alpha$ is shown for the inputs given in
eqn.~(\ref{toy-numbers}) where the true value of $\alpha=75^\circ$.  The
$B^0\to \rho/\omega~\pi^0$ results are shown as a dashed curve, the
$B^0\to \rho/\omega~\eta$ results are shown as a dotted curve while the
sum is shown as a solid curve. The magnitude of the penguin for the
minimum $\chi^2$ solution is shown in units of $10^{-3}$ by the
dot-dashed curve.
}
\vspace*{0.2 in}
\epsfxsize 3.5 in
\mbox{\epsfbox{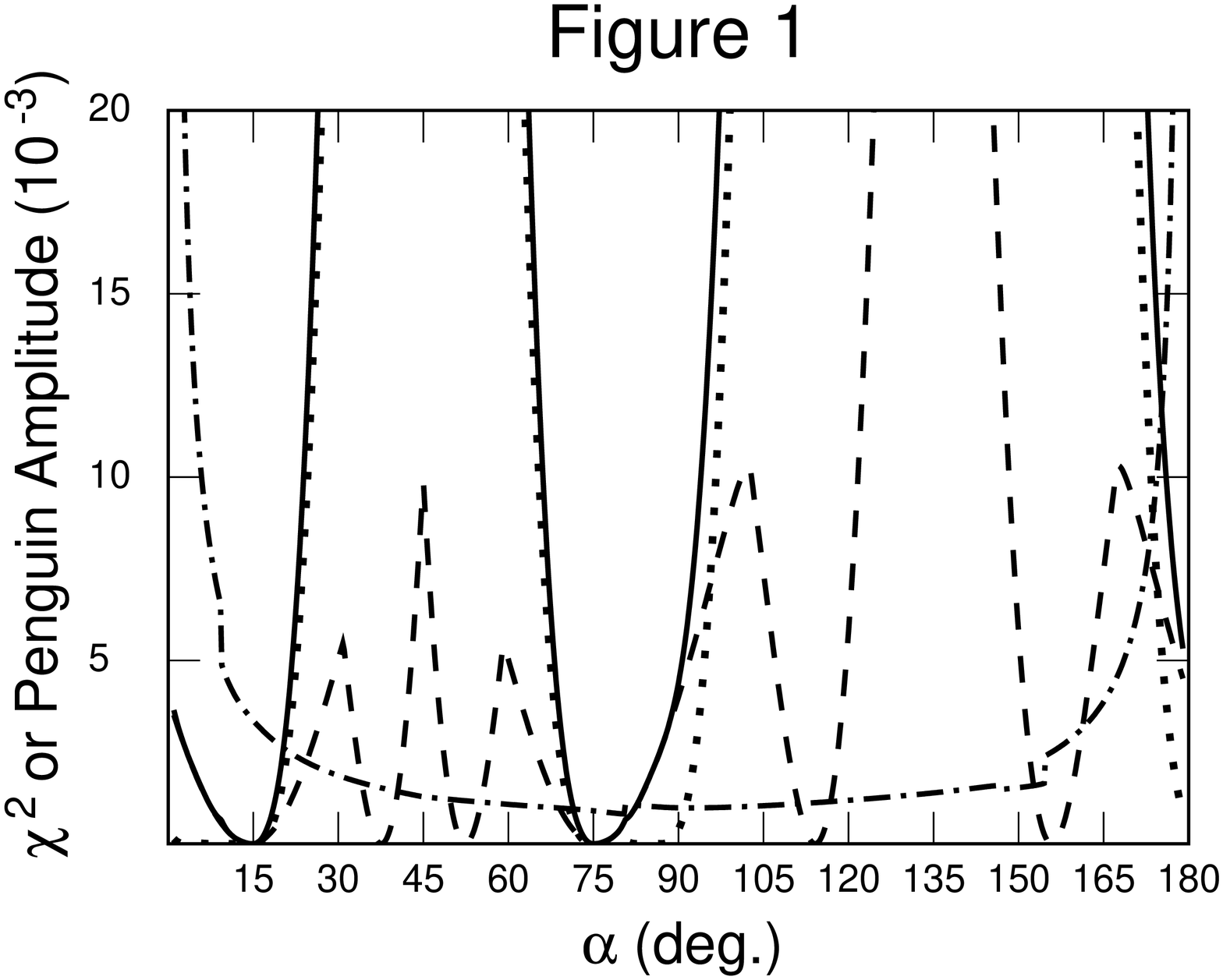}}
\end{figure}

\begin{figure}
\caption{
The $\chi^2$ function and penguin amplitude as in Fig.~1 with
$\alpha=75^\circ$ and the 
inputs as in 
eqn.~(\ref{toy-numbers}) except with $\psi^\eta_p=190^\circ$.
}
\vspace*{0.2 in}
\epsfxsize 3.5 in
\mbox{\epsfbox{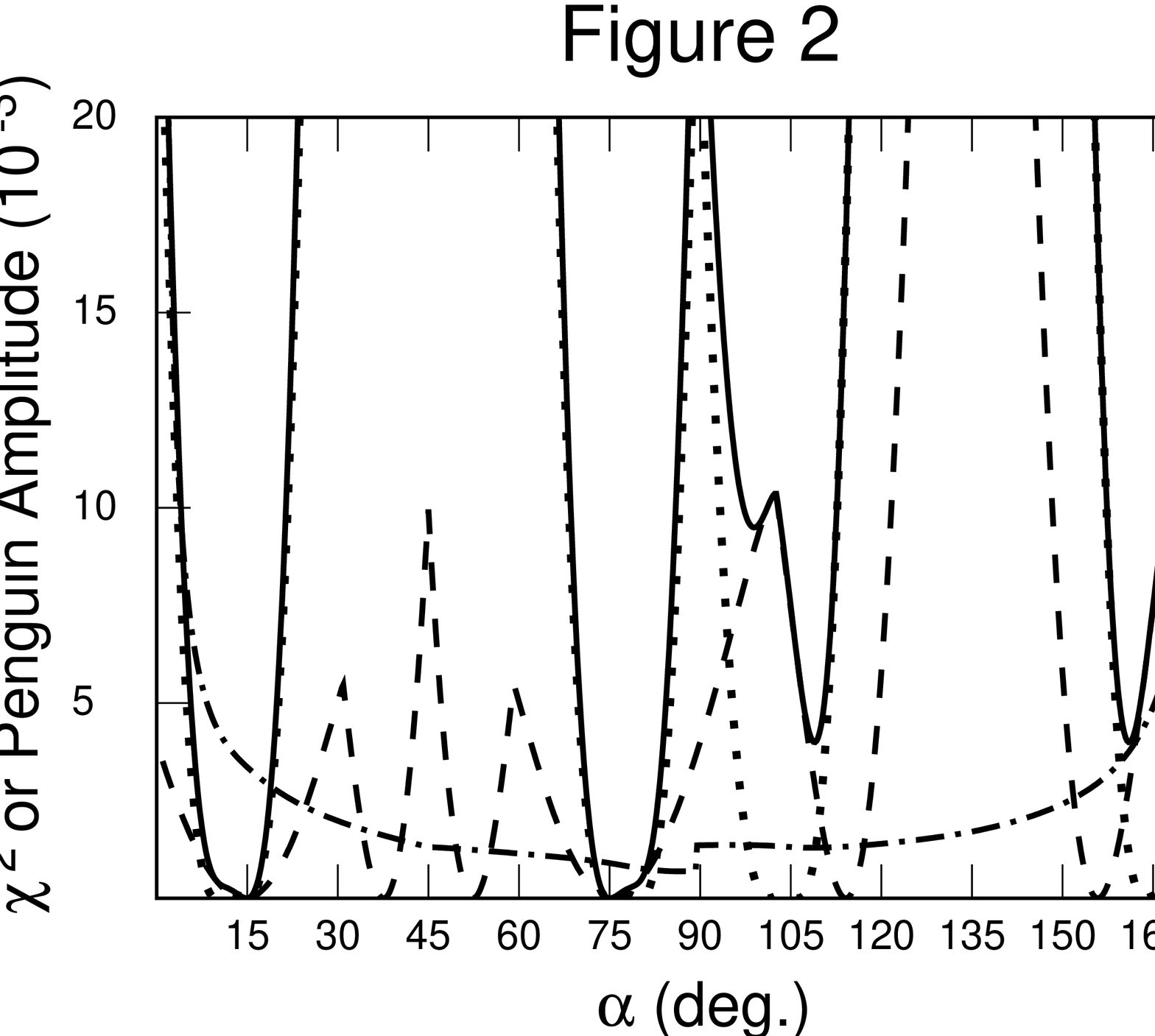}}
\end{figure}


\begin{thebibliography}{99}
\bibitem{etanote} Note that the relatively large production of 
$K\eta^\prime$ is not well understood. A number of
explanations of this effect have been suggested including constructive
interference between the $s\xba s$ and light quark components of the
$\eta/\eta^\prime$~\cite{lipkin}, anomalous coupling of the gluon to the
SU(3) singlet component~\cite{eta_anom} and large intrinsic charm
contributions~\cite{intrinsic_c}. In all of these proposed explanations,
this decay still has the CKM character of a penguin and so the discussion
here is not effected.
\bibitem{kaon_note} The only difference between the $\x=K_S$ and the
$\x=K_L$ data sets is the sign of $Z^I$. Taking this into account, the two
data sets can be combined.
\end{thebibliography}
\end{document}